\documentstyle[11pt]{article}

\begin{document}

\title{Using Rapidity Gaps to Distinguish Between 
Higgs Production by W and Gluon Fusion}

\author{Thais L. Lungov \thanks{ E-mail:
thais@uspif.usp.if.br (internet) or 47602::THAIS (decnet).}~~~~and~~ Carlos 
O. Escobar \thanks{ E-mail: escobar@uspif.usp.if.br 
(internet) or 47602::ESCOBAR (decnet).} \\
{\em  Instituto de F\'{\i}sica da Universidade de S\~ao Paulo} \\ 
{\em Caixa Postal 66318, 05389-970 S\~ao Paulo, Brazil.} }

\maketitle

\hfuzz=25pt

\begin{abstract}

The possibility of distinguishing between two higgs                  
production mechanisms, W
fusion and gluon fusion, due to rapidity gap existence
is investigated using the Monte Carlo event
generator PYTHIA.  It is shown that, considering the designed CM
energy and luminosity for the LHC, it is not possible to distinguish 
between the two higgs production processes as, for a given 
integrated luminosity, they lead to the same number of events 
containing a rapidity gap. 

\end{abstract}

\clearpage

\section{Introduction}
                                                                      
A rapidity gap is a region in rapidity space with no hadrons. 
It has been studied for a long time in  diffractive scattering 
physics.

In 1986 Dokshitzer, Khoze and Troyan~\cite{dok} suggested that
due to its peculiar color flux, a W fusion  process producing a Higgs 
boson could also lead to a rapidity gap. As each hadron is a color 
singlet,
when they emit a W boson, which is a color singlet too, they remain in 
singlet states, although separated in systems of quark and diquark.
Therefore, the initial hadrons do not need to exchange color.
The quark which emitted the W may exchange color with the diquark 
belonging to the same hadron, in order to fragment. 
According to the LUND  string model~\cite{string},
a string will be stretched between the quark and the diquark of each 
hadron (see fig.\ref{fig-fusao}). When both strings fragment, 
almost all the
hadrons formed are expected to be comprised, roughly speaking, 
between the incident beam
direction and that given by the jet containing the quark that 
emitted the
massive vector boson. Fig.~\ref{gap} shows a typical plot in
$\eta \times \Phi$ (rapidity $\times$ azimuthal angle) space, 
for an event of this kind~\cite{bjo}. 
The jets shown in the figure come from the hadronization of the
quark which emitted the vector boson (it is possible to use them as 
tagging jets). Each jet is supposed to occupy about a 0.7 radius 
circle~\cite{bjo}
(in $\eta \times \Phi$ space). The idea is to look for a gap 
in the region 
between the tangents to those jets. Imposing the higgs to decay 
into a Z
pair, which in turn is forced to decay into muon pairs, no  hadron
production occurs besides that from the hadronization of the beam 
remnants.
On the other hand, in gluon fusion each hadron emits a color 
octet (the
gluon), and therefore turns into a colored object. Hence the 
initial 
hadrons remnants must exchange color with each other in order 
to become
color neutral again (see fig.~\ref{fig-fusao}). Moreover, 
unlikely vector bosons,
gluons themselves emit quarks and other gluons, which will 
hadronize later,
thus filling the central rapidity region.

%%%
The purpose of this work is to discuss the possibility of 
using the existence 
of the rapidity
gap for  distinguishing Higgs boson production by W fusion from 
that  by gluon fusion. It is organized in the following way: in 
section~2,
some problems which appear in a gap analysis are discussed, and 
the approaches
already existing on this subject are presented; in section~3  the event
generation for this work is described; in section~4, the generated
events are analyzed, 
the gap survival probability and the number of events having gaps 
occurring for a fixed integrated luminosity are obtained for 
various higgs 
mass, and many situations, where different cuts had been imposed; 
it is shown that although for some cases there is a big  gap survival 
probability, for higgs produced by W fusion the number of events 
having a
gap which occur in an accelerator like the LHC is very small
for low higgs masses and null for the bigger ones;
when the number of events having gaps is bigger for W fusion, 
it is of the same order for gluon
fusion, thus being impossible to distinguish the two higgs production
processes, independently of some earlier conjectured problems, 
like pile up events. Finally,
in section~5, some ideas and conclusions are presented.
%%%

\section{Gap Survival Probability}

In order to use a rapidity gap for identifying the higgs 
production mechanism,
some problems have to be overcome.
For example, although in  a W fusion
the hadrons are expected to be close to the initial beams 
directions, it might
occur that some of them appear in the central region (in fact, 
it will be
shown that this is the case in many events). Furthermore, each 
incident proton is composed of several partons, and it is 
possible that more than one scattering occurs, therefore filling 
up the region between the tagging
jets (these are called {\it multiple interactions} in this paper).
Moreover, problems arise from the high luminosity designed 
for the accelerator
(as is the case for the LHC). More than one proton-proton 
scattering may occur 
in the same bunch crossing, producing the so-called 
{\it pile up} events.
That will produce even more hadrons, probably filling 
completely the 
central region. It will be shown, however,  that at least for 
LHC pile up
events are  not the worst problem. It is hard to distinguish  
between the two higgs production mechanisms, even when no 
pile up events are considered.
Besides those problems, there are others that will not be
discussed here such as gaps produced by statistical fluctuations 
in the background events,
gaps produced by other color singlet scattering (such as $WW 
\rightarrow Z \rightarrow ZZ$) and the most 
common ones,  diffractive scatterings.
These problems have already been studied by many 
authors~\cite{pumplin,duca,stirling,natale} and do not address 
the problem
investigated in this paper, which is the use of rapidity gaps to 
distinguish
between higgs production by gluon and W fusion.

Due to the difficulties mentioned above,  added to the smallness 
of the involved cross sections it is necessary to quantify the 
probability that a gap be observed, and furthermore, the possibility 
of using
its existence in experimental analysis. Bjorken~\cite{bjo} proposed a
variable, the {\it gap survival probability}, determined in the 
following
way: if $P(s,b)$ is the probability that two protons pass through 
each other
with impact parameter $b$ with no interaction occurring, except the 
hard one, the gap survival probability is given by:

\begin{equation}
\label{probsob}
S= \frac{\displaystyle \int F(b) P(s,b) d^2b}{\displaystyle
\int F(b) d^2b}
\end{equation}

\noindent
where $F(b)$ is a factor associated with the hard collision, being 
essentially
a measure of the overlap of the parton densities in the colliding 
hadrons. Eq.~\ref{probsob}  is evaluated under some conditions
(eikonal approximation, gaussian form for $F(b)$ and for the eikonal 
itself);
$S \sim 5~\%$ is obtained. 

In Ref.~\cite{levin} the same 
calculation is  performed, but using  several different 
models for hadron collisions. They obtain, for LHC energies, 
a gap survival probability 
lying between 5.3~\% and 22.1~\% (five models are analyzed, and just   
for  one of them, the Reggeon Model, one gets a probability as high 
as 22.1~\%; for the other four, it is below 8.2~\%).

An important observation must be made at this point. The gap survival 
probability, as proposed by Bjorken, and used in Ref.~\cite{levin}, 
gives the probability that, in a proton--proton collision, 
only one parton--parton
scattering occurs. It does not imply, necessarily, that a 
rapidity gap  exists, because an eventual gap can be filled 
by the hadronization of
the beam remnants, as already mentioned. This aspect has not 
been taken into account in the above  references.

Previous papers~\cite{sjo,fel} had used Monte Carlo 
simulation to analyze some
aspects of the problem. The main conclusions are: 
a) for W fusion  a rapidity 
distribution shows  two peaks; the dip between these peaks 
increases for 
increasing higgs masses and the distance between the 
peaks increases with CM 
energy~\cite{sjo}; when multiple interactions are included, 
a $p_t=2$~GeV
cut on charged particles recovers the dip~\cite{sjo}; 
b) $S \sim 3~\%$
is obtained for W fusion, $S> \sim 0.01~\%$ for the background 
($q \bar{q} \rightarrow WW$ and $t \bar{t} \rightarrow WW$)
and a null gap survival probability for $gg \rightarrow h$.
The present work broadens the scope of the former papers, taking 
into  account a larger range of higgs masses, and mainly, 
using the processes  cross-sections and the  LHC luminosity in 
order to obtain the number
of events per year presenting  a surviving rapidity gap instead 
of the probability of having such a gap.

\section{Event Generation}

The events for this work have been generated with PYTHIA~\cite{py}, 
using as distribution functions CTEQ set L2~\cite{cteq}. The top 
quark is supposed to have mass 174~GeV.
A simple calorimeter is simulated with LUCELL, a jet algorithm  
included in PYTHIA. That calorimeter covers the rapidity region 
from $\eta = -5$ to  $\eta=5$, with segmentation: $ \eta \times 
\Phi = \frac{\displaystyle 10} {\displaystyle 50} \times 
\frac{\displaystyle 2 \pi}{\displaystyle 30} \simeq
0.2 \times 0.2$. PYTHIA includes some models for simulating multiple 
interactions, and
we had chosen the default, that is the simplest one. This model is 
described both in the program manual and in Ref.~\cite{87}.

We have considered higgs produced both by gluon fusion and 
by W fusion, with 
mass varying from $m_{h}=300~GeV$ to $m_{h}=700~GeV$, supposing a pp 
collision with 14~TeV of CM energy. In both processes  the higgs
decay into a Z bosons pair, each of which then decay into a 
muon pair. That choice helps preventing the production of  hadrons 
that could fill an eventual  gap~\footnote{Gluon fusion, $ gg 
\rightarrow h \rightarrow ZZ \rightarrow \mu \mu \mu \mu$,
has been produced with Pyhtia's process number 102, while W fusion, 
$ gg \rightarrow h \rightarrow ZZ \rightarrow \mu \mu \mu \mu$, 
has been  generated with Pythia's process number 124.~\cite{87}}.
For both processes, three groups of events from now on called Group~I, 
Group~II and Group~III have been produced. 
There is a set of common cuts, cuts A, applied to the three groups, 
described below. 
The three groups have been submitted to different cuts (besides cuts A)
in order to 
determine the fraction  of events containing gaps using different 
selection criteria for events and gap definitions.
All these cuts have been largely  discussed in literature, and 
therefore  they will be  presented here without further justification. 
For each group the number of generated events is such that, after 
imposing the respective group cuts (without  including cuts A), 
10,000 events remain. 
The groups are defined in the following way:
a) {\bf Group I  } A tag is applied to two jets, and the gap is looked 
for between these jets. It is known that although this double tagging 
eliminates considerably the background, it reduces the signal 
too. Nevertheless
we adopted such a cut because it could enhance the rapidity gap 
signature. Here we demand that the event has at least two jets 
with $E_{\perp}>40$~GeV and $|\eta|>2$. These choices are imposed 
because the
quarks that emit the W boson acquires transversal momentum of order 
$ m_W / 2 $
and follows approximately the initial beam direction. If more than two
jets satisfy the above conditions, the two with the largest transversal 
moments are picked up.
The rapidity gap width is looked for in a region defined  
as being the distance in rapidity space
between the tagging jets, $\Delta \eta= \eta_1-\eta_2-1.4$, as seen in 
Fig.~\ref{gap}, where $\eta_1$ and $\eta_2$ are the jets rapidities. The
value 1.4 is subtracted due to jet width in rapidity space. Cuts A are 
applied too.
b){\bf Group II  } Here only one jet is tagged and the applied 
cuts are  similar to those used in Ref.~\cite{col}. The event 
is accepted if there is at least a jet with energy $E>1$~TeV 
and $2.0<|\eta|<5.0$. In this case a gap
is looked for inside a fixed interval, symmetric around $\eta=0$, 
and with 
width $\Delta \eta=4$ ($-2.0<\eta<2.0$). Cuts A are applied too.
c){\bf Group III  } No cut beyond that from cuts A have been 
applied to this Group. GEM~\cite{gem} adopted this kind of 
analysis. The gap
widths here is defined in the same way as for  Group~II.

Cuts A are applied to all events. They consist of~\cite{gem,col}:
a) $|\eta^l|<2.5$ and $ p_{\perp}^l > 10$~GeV. Most of the papers on
this matter demand four leptons obeying 
these conditions. But, as this is  
very restrictive, we have relaxed it, 
demanding four leptons with $p_{\perp}^l>10$~GeV but just 
three of them had to have $|\eta^l|<2.5$;
b) For  the signal, leptons are 
produced isolated. They were accepted if
inside a region of radius $R=\sqrt{\Phi^2+\eta^2}=0.3$ around each of
them no more than 5~GeV of transversal energy had been deposited; 
c) It must be possible to produce, using all four accepted leptons, 
two pairs with invariant mass next to the Z  mass: 
$|M_{ll}-M_Z|<10$~GeV; d) Muon identification and track matching;
e) At least one Z for which $p_Z>\sqrt{M_{zz}^2-4M_Z^2}$, 
where $M_{ZZ}$ is the Z pair invariant mass.

\section{Analysis}

\subsection{Gap Survival Probability}

The figures presented in this sections have been 
obtained in the following way:
a) 10,000 W fusion and 10,000 gluon fusion for each of the  
five higgs masses considered have been subjected  separately 
to cuts A; the fraction
of events surviving  the cuts in each case $F_{cc}(m_h,process)$ 
were then  obtained. This procedure does not affect the sample,
because the cuts applied concern   the part of the event that will not
be used in the final analysis, i.e., the leptons and Z's. 
It is not relevant which 
of the events are  thrown away in this case.
$F_{cc}$ is about 60~\% for gluon fusion and about 
70~\%  ($m_h=300$~GeV) to 80~\%  ($m_h=700$~GeV) for W fusion events;
b) For each group events, I, II and III, a number of events is 
generated such that, after being applied the specific cuts, 
10,000 events remain.
c) $N_{gap}$ is the number of events, for each Group, for each process 
and for each higgs 
mass, which survive, that is, which maintain the 
region where the gap is
searched for with no hadrons.
$N_{gap}$ is obtained for three different cases:
i) All charged particles are included, 
except for the  muons selected by cuts A;
ii) Charged particles with $p_{\perp}>1$~GeV, 
except for the  muons selected by cuts A;
iii) Charged particles with $p_{\perp}>2$~GeV, 
except for the  muons selected by cuts A.
c) $N_{gap}$ is divided by the {\bf total } number of generated 
events (before any kind of cut is done), 
producing $S_{bef}$ for the three 
situations analyzed above, (i), (ii) and (iii).
d) For each case, $S_{bef}$ is  multiplied by the corresponding
fraction of events surviving to cuts A and by 100, producing $S$.

Next figures represent $S$.
A) {\bf Fig.~\ref{a71-72}} shows the gap survival 
probability for events 
from Group~I. In the upper part, multiple interactions 
have not been added yet. 
If no $p_{\perp}$ cut is applied (Fig.~\ref{a71-72}.a), 
$S$ is very small for any higgs 
mass, unlike what could be expected 
from theoretical approaches.  It occurs 
probably because hadrons produced by the fragmentation of the 
beam remnants  reach
the detector central region in many events. 
This fact shows that even if
the main interaction could occur separated 
from the secondary ones,
few events would be completely clean.  
When $p_{\perp}$ cuts are considered, however, 
the gap survival probability
increases, as may be seen in Fig~\ref{a71-72}~(b) 
and (c), and if just 
$p_{\perp} > 2$~GeV particles are accepted, $S$ 
lies between 5 and 9\%
for higgs masses between 300 and 700~GeV. $S$ grows with higgs mass
as later observed~\cite{dok}.
Things change completely when multiple interactions are included.
Good results are obtained just when 
particles with $p_{\perp}<2$~GeV are left behind (Fig.~\ref{a71-72}). 
If no $p_{\perp}$ cut 
is imposed, no gap is found for any higgs mass and any process.
B) In Fig.~\ref{c74-75} the same analysis is performed for 
Group~II events. In such case, even without  $p_{\perp}$ cuts, 
$S$ lies between 4 and 8\%. Using $p_{\perp}$ cuts, results are even 
better for W fusions, but some of the gluon fusions 
will have rapidity gaps
too. It should  be noted that unlike for W fusions, for gluon fusion, 
$S$ decreases with the higgs mass.
Taking into account multiple interactions,  
it is again clear that $p_{\perp}$ cuts 
have to be imposed; if only particles with $p_{\perp}>1$~GeV 
are accepted,  $S$ lies 
between 3 and 5\% for W fusion. Nevertheless, some gluon fusion events
will produce gaps too, mainly for lower higgs masses. 
If $p_{\perp}>2$~GeV
is imposed, a more expressive $S$ value is found for W fusion, 
between 22\% and 33\%. But once more, the same cut leads gluon fusion to
produce events with gaps.
C) Fig.~\ref{e78-710} shows the same analysis for Group~III events. 
Once more the gap survival probability increases when $p_{\perp}$
cuts are applied, both for W and gluon fusion.  
The results here are slightly
smaller than for Group~II.
Group~III events (Fig.~\ref{e78-710}) 
show results quite similar to that from 
Group~II. If no $p_{\perp}$ cuts are applied, $S \sim 0$, and for 
increasing $p_{\perp}$ cuts, $S$ grows up both for W and gluon fusion.
Nevertheless, $S$ behaves oppositely with increasing higgs mass in W 
fusion and in gluon fusion.

Based upon what had been seen until now, one could conclude that in some 
circumstances the gap presence is very clear. For example, for a heavy
higgs $m_h \sim 700$~GeV, $S \sim 26$~\% for Group~III events
and $S \sim 32$~\% for Group~II events, in both cases taking into
account final charged hadrons  with $p_{\perp}>2$~GeV, $\Delta \eta=4$
and with multiple interactions included.  In both cases, for gluon 
fusion $S < 5$~\%. For a lighter higgs, the results are not so good. 
For $m_h=300$~GeV, $S \sim 15\%$ for Group~III events and $S \sim 
23\%$ for Group~II events, but the respective $S$ values for 
gluon fusion are 6\% and 8\%. As all those events passed by the same 
cuts, one could have  events with a rapidity gap that could had 
been produced either by gluon or  W fusion.

\subsection{Number of Events/Year}

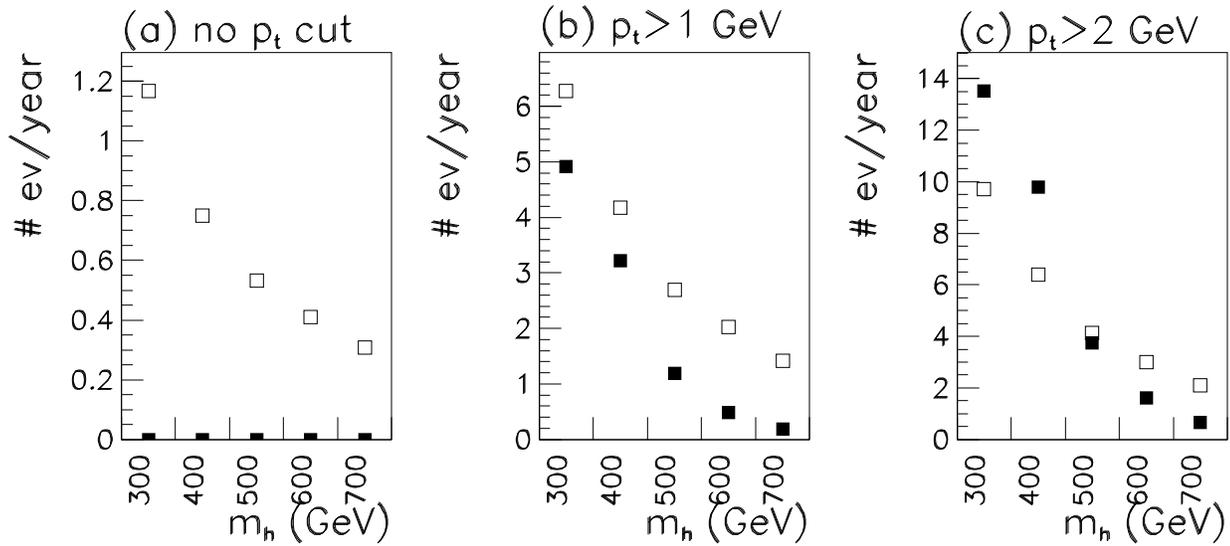
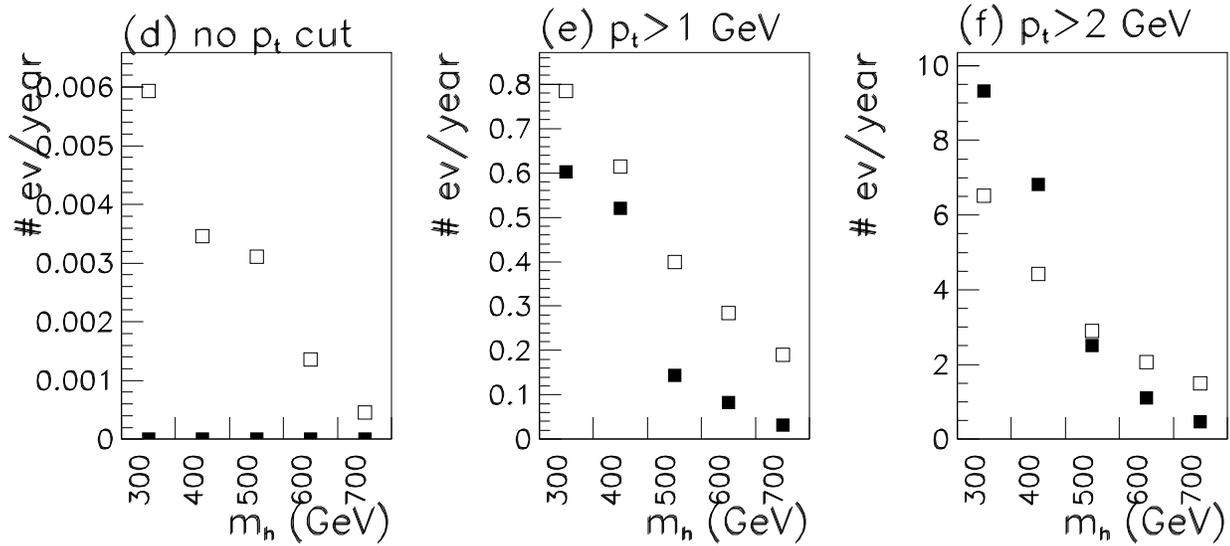
\begin{figure}[p]
\caption{Number of events having a gap  for GROUP~III events.}
\label{g3eta4ea}
\end{figure}

But there is one very important point 
that should be included in the analysis.
It is the cross section for each of the processes, gluon and W fusion 
producing higgs. To get the next figures, $S$ has been multiplied by the
respective process cross section, and by the luminosity LHC is supposed
to have ($ {\cal L}=10^{34} \hbox{cm}^{-2} \hbox{s}^{-1} \sim  
100~\hbox{events/fb-year}$). When these factors are taken into account, 
Fig.~\ref{g1ea}, \ref{g2eta4ea} and
\ref{g3eta4ea} show the number of 
events which will keep a gap in an year for $m_h$
in the range 300-700~GeV. They are obtained, respectively from Group~I, 
Group~II with $\Delta \eta=4$ and Group~III with $\Delta \eta=4$ events.
Each figure presents in the upper part, the results obtained without 
including multiple interactions and 
in the lower part, results including 
multiple interactions. As before, for both situations, 
three cases have been considered: a) all charged hadrons have 
been taken into account; b) charged
hadrons with $p_{\perp}>1$~GeV taken into account; c) charged
hadrons with $p_{\perp}>2$~GeV taken into account.

It is not difficult to see that the results are 
not very good, either including
or not multiple interactions. For Group~I (Fig.~\ref{g1ea}), less than 
two events with a higgs produced by W fusion will produce a gap in each 
year, when final charged hadrons with 
$p_{\perp}>2$~GeV are counted. With a 
softer cut, not even one event will be observed in one year.
For Group~II, although the number of events 
with rapidity gap produced by W
fusion processes is larger than that for Group~I, 
it is still small and,
what is worse, has the same magnitude that gluon fusion process has. 
For Group~III, the only situation in which event with 
rapidity gap could 
be expected is that showed in Fig.~\ref{g3eta4ea}(a), 
which is not a realistic one,
since no multiple interaction has been included.

\section{Conclusion}

Some conclusions may be drawn from our investigation: 
a)for W fusion, $S>$ increases  with
$m_h$, and for gluon fusion this behavior is opposite; b) on the 
other hand, $N_{ev/year}$ decreases with 
$m_h$, both for gluon and W fusion; c)no gap could be 
observed without $p_t$ cuts; d) after $p_t$ cuts, both W 
and gluon fusion have gaps, therefore being 
impossible to use the gap existence in distinguishing them; 
e) the above results do not depend on pile up, which have not 
been included; f) as the integrated luminosity is the same 
for W and gluon fusion, and for any higgs 
mass, the cross section is  responsible for the $N_{ev/year}$ 
behavior with higgs mass.

\section{Acknowledgments}

This work was supported by FAPESP (TLL) and by CNPq (COE and TLL).

\begin{figure}[p]
\caption{Color  exchange for (a) a W fusion and for (b) a gluon fusion,
according to LUND string model}.
\label{fig-fusao}
\end{figure}

\begin{figure}[p]
\caption{Expected legoplot for an event from  process
$pp \rightarrow WWX \rightarrow hX \rightarrow ZZX \rightarrow \mu^{+}
\mu^{-} \mu^{+} \mu^{-}$. It shows the way $\Delta\eta$ is defined 
for Group~I events}
\label{gap}
\end{figure}

\begin{figure}[p]
\caption{$S$ for GROUP~I events.}
\label{a71-72}
\end{figure}

\begin{figure}[p]
\caption{$S$ for GROUP~II events.}
\label{c74-75}
\end{figure}

\begin{figure}[p]
\caption{$S$ for GROUP~III events.}
\label{e78-710}
\end{figure}

\begin{figure}[p]
\caption{Number of events having a gap  for GROUP~I events.}
\label{g1ea}
\end{figure}

\begin{figure}[p]
\caption{Number of events having a gap  for GROUP~II events.}
\label{g2eta4ea}
\end{figure}

%\newpage

\end{document}